\documentclass[12pt]{article}

\usepackage{amsmath,amssymb,amsfonts,amsbsy}

\usepackage{cite}

\usepackage{graphicx}

\usepackage{wrapfig}

\usepackage[font=small,labelfont=bf,labelsep=period]{caption}

\usepackage{psfrag}

\def\slashchar#1{\setbox0=\hbox{$#1$}
  \dimen0=\wd0
  \setbox1=\hbox{/} \dimen1=\wd1
  \ifdim\dimen0>\dimen1
     \rlap{\hbox to \dimen0{\hfil/\hfil}}
     #1
  \else
     \rlap{\hbox to \dimen1{\hfil$#1$\hfil}}
     /
  \fi}

\begin{document}

\addcontentsline{toc}{subsection}{{TRANSVERSITY GENERALIZED PARTON DISTRIBUTIONs from $\gamma \, N\to \pi \, \rho_T N' $   WITH A LARGE  $\pi \rho_T$ INVARIANT MASS  }\\
{\it M. El Beiyad}}


\setcounter{section}{0}

\setcounter{subsection}{0}

\setcounter{equation}{0}

\setcounter{figure}{0}

\setcounter{footnote}{0}

\setcounter{table}{0}

\begin{center}

\textbf{TRANSVERSITY GPDs FROM $\gamma \, N\to \pi \, \rho_T N' $   WITH A LARGE  $\pi \rho_T$ INVARIANT MASS}

\vspace{5mm}

\underline{M. El Beiyad}$^{\,1,2\,\dag}$,  B. Pire$^{\,1}$, M. Segond$^{\,3}$, L. Szymanowski$^{4}$  and S. Wallon$^{2,5}$

\vspace{5mm}

\begin{small}

  (1) \emph{Centre  de Physique Th{\'e}orique, \'Ecole Polytechnique, CNRS,
   91128 Palaiseau, France} \\

(2) \emph{LPT, Universit\'e d'Orsay, CNRS, 91404 Orsay, France}\\

  (3) \emph{Institut f\"ur Theoretische
Physik, Universit\"at Leipzig,  D-04009 Leipzig, Germany} \\

(4) \emph{Soltan Institute for Nuclear Studies, Warsaw, Poland}\\

(5) \emph{UPMC Univ. Paris 06, facult\'e de physique, 4 place Jussieu, 75252 Paris Cedex 05, France}

 $\dag$ \emph{E-mail: mounir@cpht.polytechnique.fr}
\end{small}

\end{center}

\vspace{0.0mm} 

\begin{abstract}
The chiral-odd transversity generalized parton distributions  of the nucleon can be accessed experimentally through the exclusive photoproduction process $\gamma + N\ \to\  \pi+\rho_T +  N'$ ,  with or without beam and target polarization, provided the vector meson is produced in a transversely polarized state.  The kinematical domain of factorization is defined through  a large invariant mass of the  meson pair and  a small transverse momentum of the final nucleon. We  calculate perturbatively the scattering amplitude at leading order in $\alpha_s$ and build a simple model for the dominant transversity GPD $H_T(x,\xi ,t)$ based on the concept of double distribution.  Counting rate estimates are in progress.\end{abstract}

\vspace{7.2mm} 

Transversity quark distributions in the nucleon remain among the most unknown leading twist hadronic observables. This is mostly due to their chiral odd character which enforces their decoupling in most hard amplitudes. After the pioneering studies \cite{tra}, much work \cite{Barone} has been devoted to the exploration of many channels but experimental difficulties have challenged the most promising ones.

On the other hand, tremendous progress has been recently witnessed on the QCD description of hard exclusive processes, in terms of generalized parton distributions (GPDs) describing the 3-dimensional content of hadrons. Access to the chiral-odd transversity GPDs~\cite{defDiehl}, noted  $H_T$, $E_T$, $\tilde{H}_T$, $\tilde{E}_T$, has however turned out to be even more challenging~\cite{DGP} than the usual transversity distributions: one photon or one meson electroproduction leading twist amplitudes are insensitive to transversity GPDs.  The strategy which we follow here, as initiated in Ref.~\cite{IPST}, is to study the leading twist contribution to processes where more mesons are present in the final state; the hard scale which allows to probe the short distance structure of the nucleon
is  $s=M_{\pi \rho}^2\, \sim |t'|$ in the fixed angle regime.
%
%
In the example developed previously~\cite{IPST}, the process under study was the high energy photo (or electro) diffractive production of two vector mesons, the hard probe being the virtual "Pomeron" exchange (and the hard scale was the virtuality of this pomeron), in analogy with the virtual photon exchange occuring in the deep electroproduction of a meson. A similar strategy has also been advocated recently in Ref.~\cite{kumano} to enlarge the number of processes which could be used to extract information on chiral-even GPDs.

The process we study here 
\begin{equation}
\gamma + N \rightarrow \pi + \rho_T + N'\,,
\label{process}
\end{equation}
is a priori sensitive to chiral-odd GPDs because of the chiral-odd character of the leading twist distribution amplitude (DA) of the transversely polarized $\rho$ meson.  The estimated rate depends of course much on the magnitude of the chiral-odd GPDs. Not much is known about them, but model calculations have been developed in~\cite{IPST,Sco,Pasq,othermodels}; moreover, a few moments have been computed on the lattice~\cite{lattice}.

To factorize the amplitude of this process we use  the now classical proof of the factorization of exclusive scattering at fixed angle and large energy~ \cite{LB}. The amplitude for the process $\gamma + \pi \rightarrow \pi + \rho $ is written as the convolution of mesonic DAs  and a hard scattering subprocess amplitude $\gamma +( q + \bar q) \rightarrow (q + \bar q) + (q + \bar q) $ with the meson states replaced by collinear quark-antiquark pairs. This is described in Fig.~1a. The absence of any pinch singularities (which is the weak point of the proof for the generic case $A+B\to C+D$) has been proven in Ref.~\cite{FSZ} for the case  of interest here.
We then extract from the factorization procedure of the deeply virtual Compton scattering amplitude near the forward region the right to replace in Fig.~1a the lower left meson DA by a $N \to N'$ GPD, and thus get Fig.~1b. We introduce
$\xi$ as the skewness parameter which can be written in terms of the meson pair squared invariant mass
$M^2_{\pi\rho}$ as
\begin{equation}
\label{skewedness}
\xi = \frac{\tau}{2-\tau} ~~~~,~~~~\tau =
\frac{M^2_{\pi\rho}}{S_{\gamma N}-M^2}\,.
\end{equation}

Indeed the same collinear factorization property bases the validity of the leading twist approximation which either replaces the meson wave function by its DA or the $N \to N'$ transition to its GPDs. A slight difference is that light cone fractions ($z, 1- z$) leaving the DA are positive, while the corresponding fractions ($x+\xi,\xi-x$) may be positive or negative in the case of the GPD. The calculation will show that this difference does not ruin the factorization property, at least at the Born order  we are studying here.

\begin{figure}[h]
\begin{center}
\psfrag{z}{\begin{small} $\hspace{-.1cm}z$ \end{small}}
\psfrag{zb}{\raisebox{-0cm}{ \begin{small}$\hspace{-.1cm}\bar{z}$\end{small}} }
\psfrag{gamma}{\raisebox{+.1cm}{ $\,\gamma$} }
\psfrag{pi}{$\,\pi$}
\psfrag{rho}{$\,\rho_T$}
\psfrag{TH}{\hspace{-0.2cm} $T_H$}
\psfrag{tp}{\raisebox{.6cm}{{\begin{small}     $t'$       \end{small}}}}
\psfrag{s}{\begin{small}$s$ \end{small}}
\psfrag{Phi}{ \hspace{-0.3cm} $\phi$}
\hspace{-0.7cm}
\raisebox{.7cm}{\includegraphics[width=7cm]{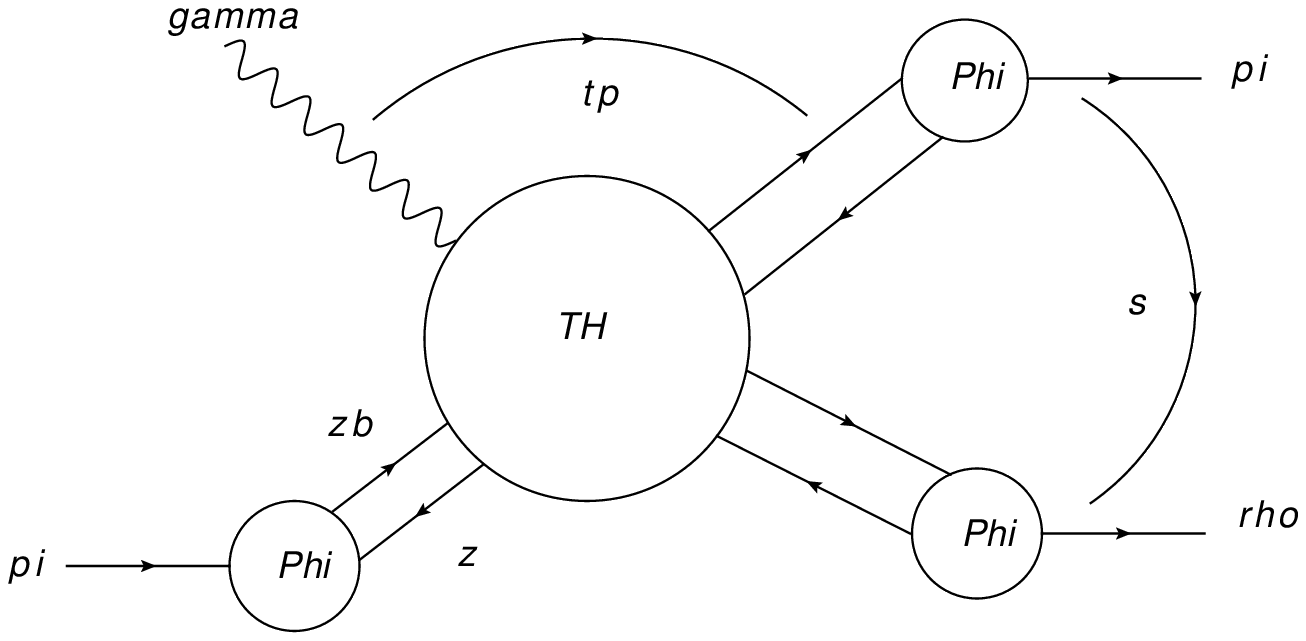}}~~\hspace{0.6cm}
\psfrag{piplus}{$\,\pi$}
\psfrag{rhoT}{$\,\rho_T$}
\psfrag{M}{\hspace{-0.3cm} \begin{small} $M^2_{\pi \rho}$ \end{small}}
\psfrag{x1}{\hspace{-0.7cm} \begin{small}  $x+\xi $  \end{small}}
\psfrag{x2}{ \hspace{-0.2cm}\begin{small}  $x-\xi $ \end{small}}
\psfrag{N}{ \hspace{-0.4cm} $N$}
\psfrag{GPD}{ \hspace{-0.6cm}  $GPDs$}
\psfrag{Np}{$N'$}
\psfrag{t}{ \raisebox{-.1cm}{ \hspace{-0.5cm} \begin{small}  $t$  \end{small} }}
\psfrag{tp}{\raisebox{.5cm}{{\begin{small}     $t'$       \end{small}}}}
 \includegraphics[width=7cm]{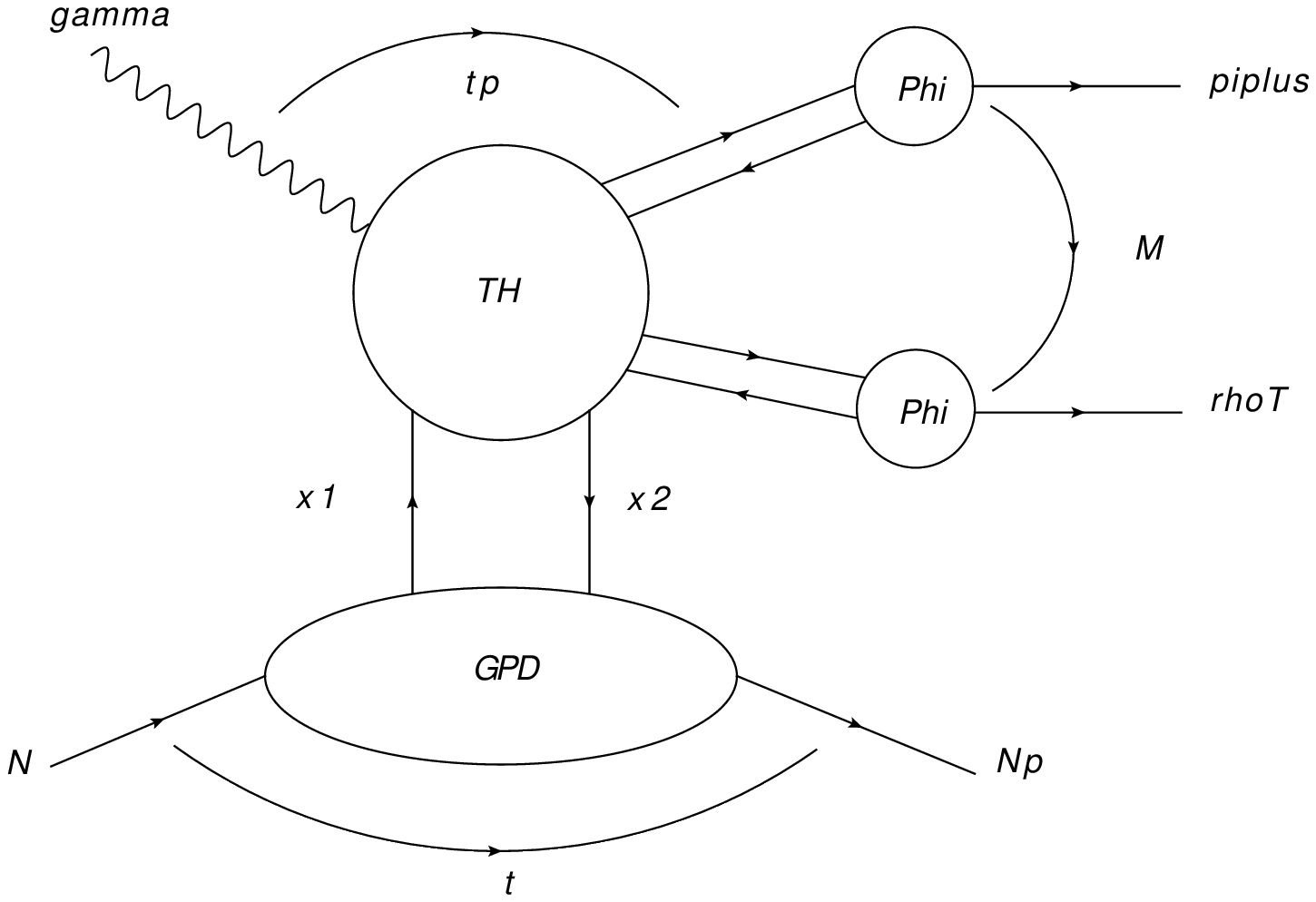}
\caption{a) (left) Factorization of the amplitude for the process $\gamma + \pi \rightarrow \pi + \rho $ at large $s$ and fixed angle (i.e. fixed ratio $t'/s$); b) (right) replacing one DA by a GPD leads to the factorization of the amplitude  for $\gamma + N \rightarrow \pi + \rho +N'$ at large $M_{\pi\rho}^2$.}
\label{feyndiag}
\end{center}
\end{figure}

In order for the factorization of a partonic amplitude to be valid, and the
leading twist calculation to be sufficient, one should avoid the dangerous
kinematical regions where a small momentum transfer is exchanged in the
upper blob, namely small $t' =(p_\pi -p_\gamma)^2$ or small
$u'=(p_\rho-p_\gamma)^2$, and the regions where strong interactions between
two hadrons in the final state are non-perturbative, namely where one of the
invariant squared masses $(p_\pi +p_{N'})^2 , (p_\rho +p_{N'})^2, (p_\pi +p_\rho)^2$
is in the resonance region. 

The scattering amplitude of the process (\ref{process}) is written in the factorized form :
\begin{equation}
\label{AmplitudeFactorized}
\mathcal{A}(t,M^2_{\pi\rho},p_T)  = \int_{-1}^1dx\int_0^1dv\int_0^1dz\ T^q(x,v,z) \, H^{q}_T(x,\xi,t)\Phi_\pi(z)\Phi_\bot(v)\,,
\end{equation}
where
$T^q$ is the hard part of the amplitude and
the transversity GPD of a parton $q$  in the nucleon target which dominates at small momentum transfer is defined by~\cite{defDiehl}
\[
\langle N'(p_2),\lambda'|\bar{q}\left(-\frac{y}{2}\right)\sigma^{+j}\gamma^5 q \left(\frac{y}{2}\right)|N(p_1),\lambda \rangle  = \bar{u}(p',\lambda')\sigma^{+j}\gamma^5u(p,\lambda)\int_{-1}^1dx\ e^{-\frac{i}{2}x(p_1^++p_2^+)y^-}H_T^q\,,
\]
where $\lambda$ and $\lambda'$ are the light-cone helicities of the nucleon $N$ and $N'$.
The chiral-odd  DA for the transversely polarized meson vector $\rho_T$,  is defined, in leading twist 2, by the matrix element~\cite{defrho}
\[
\langle 0|\bar{u}(0)\sigma^{\mu\nu}u(x)|\rho^0_T(p,\epsilon_\pm) \rangle =\frac{i}{\sqrt{2}} (\epsilon^\mu_{\pm}(p) p^\nu - \epsilon^\nu_{\pm}(p) p^\mu)f_\rho^\bot\int_0^1du\ e^{-iup\cdot x}\ \phi_\bot(u)\,,
\]
where $\epsilon^\mu_{\pm}(p_\rho)$ is the $\rho$-meson transverse polarization and with $f_\rho^\bot$ = 160 MeV.\\

\begin{figure}[!h]
\centerline{$\begin{array}{cc}
\psfrag{fpi}{$\,\phi_\pi$}
\psfrag{fro}{$\,\phi_\rho$}
\psfrag{z}{\begin{small} $z$ \end{small}}
\psfrag{zb}{\raisebox{-.2cm}{ \begin{small}$\hspace{-.3cm}-\bar{z}$\end{small}} }
\psfrag{v}{\begin{small} $v$ \end{small}}
\psfrag{vb}{\raisebox{-.1cm}{ \begin{small}$\hspace{-.4cm}-\bar{v}$\end{small}} }
\psfrag{gamma}{$\,\gamma$}
\psfrag{pi}{$\,\pi$}
\psfrag{rho}{$\,\rho_T$}
\psfrag{N}{$N$}
\psfrag{Np}{$\,N'$}
\psfrag{H}{\hspace{-0.2cm} $H_T(x,\xi,t)$}
\psfrag{p1}{\begin{small}     $p_1$       \end{small}}
\psfrag{p2}{\begin{small} $p_2$ \end{small}}
\psfrag{p1p}{\hspace{-0.8cm}  \begin{small}  $p_1'=(x+\xi) p$  \end{small}}
\psfrag{p2p}{\hspace{-0.2cm} \begin{small}  $p_2'=(x-\xi) p$ \end{small}}
\psfrag{q}{\begin{small}     $\hspace{-.5cm} q$       \end{small}}
\psfrag{ppi}{\begin{small} $p_\pi$\end{small}}
\psfrag{prho}{\begin{small} $p_\rho$\end{small}}
\includegraphics[width=7.5cm]{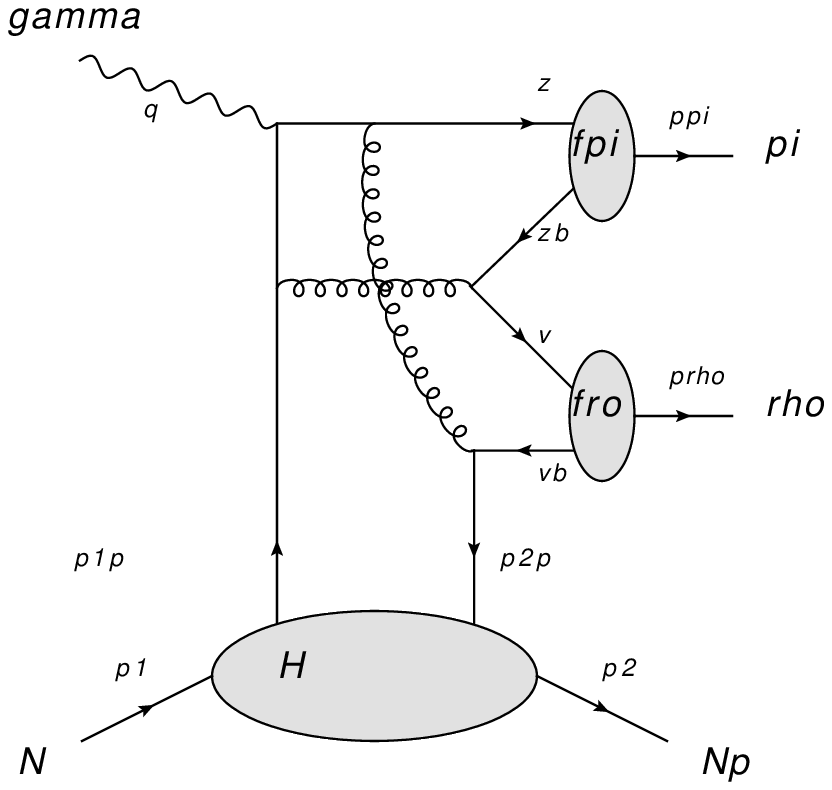}&\hspace{-0.3cm}
\psfrag{fpi}{$\,\phi_\pi$}
\psfrag{fro}{$\,\phi_\rho$}
\psfrag{z}{\begin{small} $z$ \end{small}}
\psfrag{zb}{\raisebox{-.2cm}{ \begin{small}$\hspace{-.3cm}-\bar{z}$\end{small}} }
\psfrag{v}{\begin{small} $v$ \end{small}}
\psfrag{vb}{\raisebox{-.1cm}{ \begin{small}$\hspace{-.4cm}-\bar{v}$\end{small}} }
\psfrag{gamma}{$\,\gamma$}
\psfrag{pi}{$\,\pi$}
\psfrag{rho}{$\,\rho_T$}
\psfrag{N}{$N$}
\psfrag{Np}{$\,N'$}
\psfrag{H}{\hspace{-0.2cm} $H_T(x,\xi,t)$}
\psfrag{p1}{\begin{small}     $p_1$       \end{small}}
\psfrag{p2}{\begin{small} $p_2$ \end{small}}
\psfrag{p1p}{\hspace{-0.8cm}  \begin{small}  $p_1'=(x+\xi) p$  \end{small}}
\psfrag{p2p}{\hspace{-0.2cm} \begin{small}  $p_2'=(x-\xi) p$ \end{small}}
\psfrag{q}{\begin{small}     $\hspace{-.5cm}q$       \end{small}}
\psfrag{ppi}{\begin{small} $p_\pi$\end{small}}
\psfrag{prho}{\begin{small} $p_\rho$\end{small}}
\includegraphics[width=7.5cm]{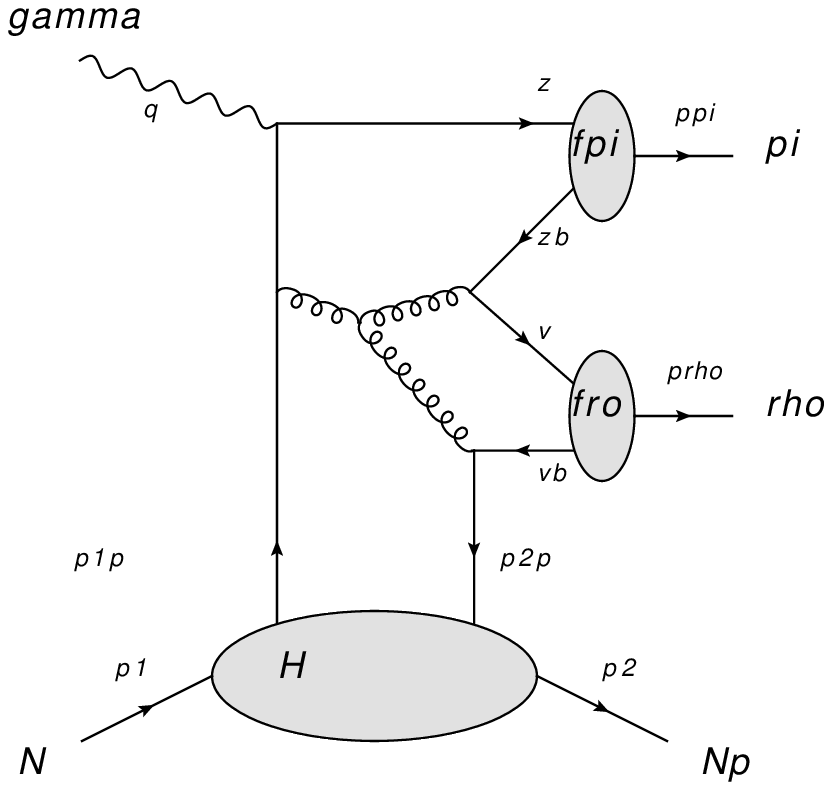}
\end{array}$}
\caption{Two representative diagrams without (left) and with (right) three gluon coupling.}
\label{feyndiageued}
\end{figure}

Two classes  of Feynman diagrams (see Fig.2), without  and  with  a 3-gluon vertex, describe this process.
In both cases, an interesting symmetry allows to deduce the contribution of some diagrams from other ones, reducing our task to the calculation of half the 62 diagrams involved in the process.
The scattering amplitude gets both a real and an imaginary parts. Integrations over $v$ and $z$ have been done analytically whereas numerical methods are used for the integration over $x$.
Various observables can be calculated with this amplitude. We stress that even the unpolarized differential cross-section $\frac{d\sigma}{dt \,du' \, dM^2_{\pi\rho}}$
is sensitive to the transversity GPD.
Rate estimates are under way, based on a double distribution model for $H_T$ . Preliminary results allow
us to
provisionnally conclude that the photoproduction of a transversely polarized vector meson on a nucleon target is a good way to reach informations on the generalized chiral-odd quark content of the proton. For instance, if the JLab 12 GeV upgrade gets the nominal luminosity  ($\mathcal{L} \sim 10^{35}$ cm$^2$. s$^{-1}$), we expect a sufficient number of events per year to extract the dominant transversity GPD.


We are grateful to Igor Anikin, Markus Diehl, Samuel Friot and Jean Philippe Lansberg for useful 
discussions.
This work is partly supported by the French-Polish scientific agreement Polonium 7294/R08/R09,  the ECO-NET program, contract 18853PJ, the ANR-06-JCJC-0084, the Polish Grant N202 249235 and the DFG (KI-623/4).





\begin{thebibliography}{99} 

\bibitem{tra}
  J.~P.~Ralston and D.~E.~Soper,
  Nucl.\ Phys.\ B {\bf 152}, 109 (1979);
  X.~Artru and M.~Mekhfi,
  Z.\ Phys.\ C {\bf 45}, 669 (1990);
  J.~L.~Cortes {\em et al.},
  Z.\ Phys.\ C {\bf 55}, 409 (1992);
  R.~L.~Jaffe and X.~D.~Ji,
  Phys.\ Rev.\ Lett.\  {\bf 67}, 552 (1991).

\bibitem{Barone} 
  V.~Barone, A.~Drago and P.~G.~Ratcliffe,
  Phys.\ Rept.\  {\bf 359}, 1 (2002);
  M.~Anselmino,
  arXiv:hep-ph/0512140;
   B.~Pire and L.~Szymanowski,
  Phys.\ Rev.\ Lett.\  {\bf 103}, 072002 (2009).


\bibitem{defDiehl}
  M.~Diehl,
  Eur.\ Phys.\ J.\ C {\bf 19}, 485 (2001).

\bibitem{DGP}  
M.~Diehl {\em et al.},
  Phys.\ Rev.\  D {\bf 59}, 034023 (1999);
  J.~C.~Collins and M.~Diehl,
  Phys.\ Rev.\  D {\bf 61}, 114015 (2000).
  
 \bibitem{IPST}   
 D.~Yu.~Ivanov {\em et al.},  
  Phys.\ Lett.\  B {\bf 550}, 65 (2002);
 R.~Enberg {\em et al.},  
  Eur.\ Phys.\ J.\  C {\bf 47}, 87 (2006).

 \bibitem{kumano}
   S.~Kumano {\em et al.},
 Phys.\ Rev.\  D {\bf 80}, 074003 (2009).
  
  \bibitem{Sco}
    S.~Scopetta,
  Phys.\ Rev.\  D {\bf 72}, 117502 (2005).
  

\bibitem{Pasq}
  M.~Pincetti, B.~Pasquini and S.~Boffi,
  Phys.\ Rev.\  D {\bf 72}, 094029 (2005) and Czech.\ J.\ Phys.\  {\bf 56}, F229 (2006).
  
  \bibitem{othermodels}
  M.~Wakamatsu,
  Phys.\ Rev.\  D {\bf 79}, 014033 (2009);
  D.~Chakrabarti {\em et al.},
  Phys.\ Rev.\  D {\bf 79}, 034006 (2009).

\bibitem{lattice}  
   M.~Gockeler {\it et al.},
   Phys.\ Rev.\ Lett.\  {\bf 98}, 222001 (2007) and
  Phys.\ Lett.\  B {\bf 627}, 113 (2005).



\bibitem{LB}
 G.~P.~Lepage and S.~J.~Brodsky,
  Phys.\ Rev.\  D {\bf 22}, 2157 (1980).
  
\bibitem{FSZ}  
 G.~R.~Farrar {\em et al.},
  Phys.\ Rev.\ Lett.\  {\bf 62}, 2229 (1989).
  
  
   \bibitem{impact}
   M.~Burkardt,
  Phys.\ Rev.\  D {\bf 62} (2000) 071503;
 J.~P.~Ralston and B.~Pire,
  Phys.\ Rev.\  D {\bf 66} (2002) 111501;
   M.~Burkardt,
  Phys.\ Rev.\  D {\bf 72}, 094020 (2005);
   M.~Diehl and Ph.~Hagler,
  Eur.\ Phys.\ J.\  C {\bf 44}, 87 (2005);
   A.~Mukherjee, D.~Chakrabarti and R.~Manohar,
 AIP Conf.\ Proc.\  {\bf 1149}, 533 (2009).


\bibitem{defrho}
  P.~Ball and V.M.~Braun,
  Phys.\ Rev.\ D {\bf 54}, 2182 (1996).
\end{thebibliography}
\end{document}